\documentstyle[twoside,fleqn,espcrc2,float]{article}

\input epsf
\epsfverbosetrue

\title{Resonant inelastic x-ray scattering as a probe of optical scale 
excitations in strongly electron-correlated systems: quasi-localized 
view}

\author{Sergei M. Butorin\\
\vspace{0.3cm}
Department of Physics, Uppsala University, 
Box 530, S-751 21 Uppsala, Sweden}

\begin{document}

\begin{abstract}

An application of resonant inelastic x-ray scattering technique for 
studying of optical scale excitations in electron-correlated materials 
is discussed.  Examples are given including data obtained for $3d$ 
transition metal, lanthanide, and actinide systems.  In some cases, 
the data are compared with the results of crystal-field multiplet and 
Anderson impurity model calculations.  Advantages of this technigue 
are pointed out, such as an ability to probe an extended multiplet 
structure of the ground state configuration, which is not fully 
accessible by other spectroscopies, an extreme sensitivity of spectral 
profiles to the chemical state of the element in question and to the 
crystal-field strength, and a great potential in probing the ground 
state character (for example, ground state $J$-mixing in rare-earths) 
due to the technique's elemental selectivity and strict selection 
rules.  Issues are addressed, such as a possible deviations from the 
linear dispersion of inelastic scattering structures, corresponding to 
charge-transfer excitations, with varying excitation energies and an 
estimation of values for model parameters, involved in the description 
of charge-transfer processes.

\end{abstract}

\floatstyle{plain}
\restylefloat{figure}

\maketitle

\section{Technique and models}

To successfully describe various physical properties of a system in 
question it is necessary to obtain knowledge about the ground state 
and low-energy excited states of this system.  For 3$d$ transition 
element, lanthanide, and actinide compounds with a partly filled $d$ 
or $f$ shell, strong correlation effects, when the dispersional part 
of $d$ or $f$ bandwidth is smaller than the on-site Coulomb 
interaction $U$ between localized electrons, break down a 
single-particle picture and an atomic-like approach to characterize 
the electronic structure of these compounds is more appropriate.

In this case a state of the system without a core hole is described in 
terms of intra-atomic neutral excitations (a multiplet structure of 
the ground state electronic configuration due to electrostatic, 
exchange, crystal field, spin-orbit interactions, etc.)  and/or 
inter-atomic charge-transfer excitations.  The latter are the result 
of electron hopping from delocalized states to a localized state and 
are treated by short-range models, such as an Anderson impurity model 
\cite{R1}, using a set of parameters.  The models are represented by 
the Hamiltonian

\begin{eqnarray}
H &=& \sum_{k,\alpha,\sigma} \varepsilon_{k\alpha}n_{k\alpha\sigma} + 
\sum_{m,\sigma} \varepsilon_{m}n_{m\sigma} \nonumber  \\
  & &+\sum_{k,\alpha,m,\sigma}(V_{k{\alpha}m}\psi^{\dag}_{m\sigma}
\psi_{k\alpha\sigma} + H.c.)  \nonumber  \\
  & &+\;U\!\sum_{m,m^{\prime},\sigma,\sigma^
{\prime}}n_{m\sigma}n_{m^{\prime}\sigma^{\prime}}.
\end{eqnarray}

Important physical quantities included in this Hamiltonian are the 
delocalized- and localized-state energies $\varepsilon_{k\alpha}$ and 
$\varepsilon_{m}$, hopping matrix element $V_{k{\alpha}m}$, and $U$.  
Here $k$, $\alpha$, $\sigma$, and $m$ denote a wave vector, an index 
of the energy level in the valence band, a spin index, and an azimutal 
quantum number, respectively.  For the description of core 
spectroscopies a further term is added to the Hamiltonian to account 
for coupling between localized electron and a core hole.  The values 
of model parameters are optimized by fitting both high-energy 
spectroscopic and low-energy transport data and then employed to 
describe the character of the ground state, different ground-state 
properties, the nature and size of the band gap in insulators 
\cite{R2}, etc.

Since the interpretation of transport measurements in these regards is 
often hampered by the presence of defects and by the importance of 
electron-lattice interactions, high-energy spectroscopies which 
directly probe the electronic degrees of freedom are often used for 
preliminary estimations of model parameters.  In these estimations, it 
is important to take into account significant configurational 
dependence of model parameters which is predicted by first-principles 
calculations \cite{R3}.  In particular, removing/adding of a valent 
$d$ or $f$ electron is expected to result in a decrease/increase in 
the value of the hybridization strength $V$ which in turn may lead to 
renormalization effects for $U$.  These effects are more pronounced 
for core-level spectroscopies.  In the presence of a core hole $V$ is 
strongly reduced since the waverfunctions become more localized.  The 
renormalization of model parameters in the final state can produce a 
significant uncertainty in estimated values of these parameters in the 
ground state.  In this situation, x-ray scattering techniques become 
very attractive because the scattering process is charge-neutral.

\begin{figure}
\centerline{\epsfxsize=7.0cm \epsfbox{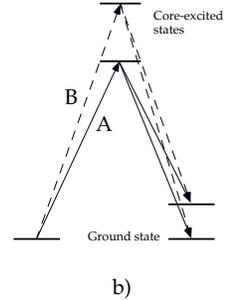}}
\caption{Schematic representation of the radiative deexcitation 
process for two different core excitations $A$ and $B$.}
\label{fig1}
\end{figure}

By now, it has been proven that at some core-thresholds of 
electron-correlated systems x-ray emission spectroscopy with 
monochromatic photon excitation can be considered as an analog of 
these techniques so that the excitation-radiative deexcitation channel 
can be treated as resonant inelastic x-ray scattering (RIXS) process.  
Final states probed via such a channel are related to eigenvalues of 
the ground state Hamiltonian.  The core-hole lifetime is not a limit 
on the resolution in this spectroscopy (see e.~g.  
Ref.~\cite{Kuiper}).  It is important to distinguish between the 
many-body description of RIXS and a single-particle approach which is 
usually applied to wide-band materials \cite{R4}.  The differences 
between two formalisms are schematically illustrated in 
Fig.~\ref{fig1}.

According to the many-body picture, an energy of a photon, scattered 
on a certain low-energy excitation, should change by the same amount 
as a change in an excitation energy of the primary beam (see a decay 
route of core excitation $B$ versus that of $A$) so that inelastic 
scattering structures have constant energy losses and follow the 
elastic peak on the emitted-photon energy scale.  In the 
single-particle view, energy positions of specific 
inelastic-scattering structures with respect to the elastic peak which 
are defined by the momentum conservation rule may vary only within the 
energy range covered by the occupied part of the valence band.  In 
Fig.~1a, this is reflected in the situation when, for core excitation 
$B$ with the higher energy, the radiative decay results in a 
transition with the lower energy than those for $A$, respectively.  In 
spite of simplifications made here, the outlined differences can be 
used to test the validity of one or another model for a system in 
question.

\begin{figure}
\centerline{\epsfxsize=8.0cm \epsfbox{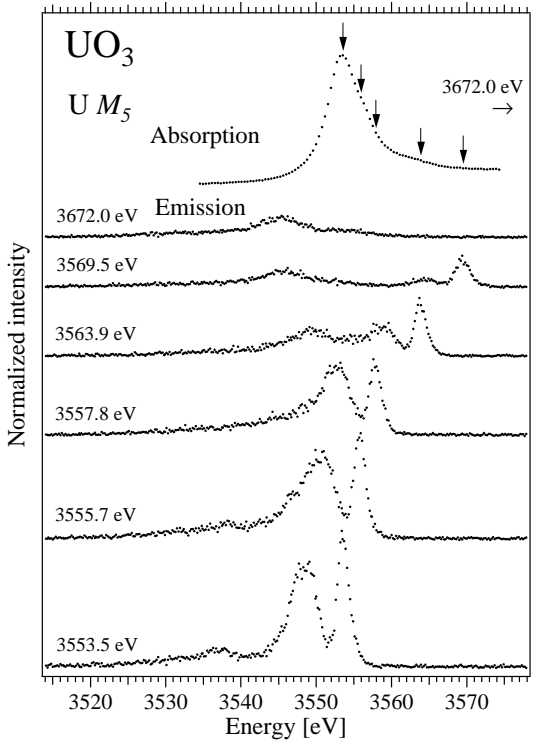}}
\caption{U $M_5$ x-ray fluorescence spectra recorded at the U 
3$d_{5/2}$ threshold of UO$_{3}$ (Ref.~\cite{R5}).  Excitation 
energies used in these measurements are indicated by arrows on the 
absorption spectrum.}
\label{fig2}
\end{figure}

As an example, we use data from Ref.~\cite{R5} (see Fig.~\ref{fig2}) 
which were obtained at the U 3$d_{5/2}$ edge of UO$_{3}$.  The 
inelastic scattering structure with the energy loss of about 5 eV is 
observed to follow the elastic peak up to 20 eV above the 3$d_{5/2}$ 
threshold while the width of the occupied part of the valence band is 
only $\sim$4 eV. This indicates the importance of electron correlation 
effects in UO$_{3}$.

Although, the information provided by the RIXS technique is similar to 
that obtained from optical absorption or low-energy 
electron-energy-loss (EELS) spectroscopies, there are some advantages 
in using this method:
\begin{enumerate}
\item the technique is not surface-sensitive, helping to avoid the 
confusion with a formation of additional states because of surface 
defects;

\item its element-specificity enables one to study of even very dilute 
compounds since metal states can be probed separately from ligand 
states;

\item the cross-section for inelastic x-ray scattering 
is strongly enhanced on the resonance in contrast to weak 
dipole-forbidden transitions in optical absorption spectra; 

\item the dipole nature of radiative transitions makes it easier to 
calculate RIXS intensities compared to $d$-$d$ ($f$-$f$) intensities 
in optical spectroscopy or in EELS.
\end{enumerate}

In calculations of resonant x-ray scattering as a second order optical 
process, only a resonant term of the modified Kramers-Heisenberg 
equation is usually used, where the spectral intensity is given by

\begin{eqnarray}
I_{qq^\prime}(\Omega,\omega) &=& \sum_{f} \Bigl| \sum_{i} { \langle f 
|D^{(1)}_{q^\prime}| i \rangle \langle i |D^{(1)}_{q}| g \rangle \over 
E_{g}+{\Omega}-E_{i}-i \Gamma_{i} / 2 } \Bigr|^2 \nonumber  \\
                             & & \times \delta (E_{g}+{\Omega}-E_{f}-
{\omega}).
\end{eqnarray}
Here, $| g \rangle$, $| i \rangle$, and $| f \rangle$ are the ground, 
intermediate, and final states with energies $E_{g}$, $E_{i}$, and 
$E_{f}$, respectively, while $\Omega$ and $\omega$ represent energies 
of incident and scattered photons, respectively.  $D$ is the dipole 
operator, $\Gamma$ stands for the intermediate state lifetime and $q$ 
and $q^\prime$ are polarizations of the light with respect to the 
quantization axis.

For the $q$-polarized incident photons, spectra detected in different 
directions with respect to the quantization axis can be described 
\cite{R6} as

\begin{eqnarray}
	I^{iso} &=& I_{q0} + I_{q1} + I_{q-1}, \nonumber  \\
    I^{90^\circ} &=& I_{q0} + {1 \over 2}(I_{q1} + I_{q-1}),   \\
    I^{0^\circ} &=& I_{q1} + I_{q-1}. \nonumber
\end{eqnarray}

For the case of studying the multiplet structure of the ground state 
configuration, the ability of resonant x-ray emission spectroscopy to 
probe low-energy excitations was first discussed by Tanaka and Kotani 
\cite{R7} in the description of resonant Cu $3d\rightarrow2p$ spectra 
of La$_2$CuO$_4$ and CuO. The difference in $d$-$d$ excitation 
profiles for $3d^9$ multiplets between these two oxides was predicted.  
However, no experimental data were available with the energy 
resolution being sufficient enough to support conclusions made by 
authors.  The first experiment which unambiguously confirmed the 
ability of this resonant technique to probe elementary excitations was 
performed on MnO \cite{R8} (two years later, high-resolution RIXS data 
at the Cu $3p$ resonance of cuprates were published \cite{Kuiper} 
which are in good agreement with theoretical predictions).  Prior to 
this, probing of $f$-$f$ excitations in rare-earths was discussed in 
Ref.~\cite{R9}.  The efficiency of the technique in studies of 
charge-transfer excitations for valence electrons in correlated 
systems was first demonstrated by Butorin {\it et al.} \cite{R5} for 
both soft and intermediate-energy x-ray regions.

Present paper is to large extent centered on the description of RIXS 
as a tool for studying of elementary excitations.  Various aspects of 
probing the charge-transfer excitations by this spectroscopy and data 
interpretation within the Anderson impurity model framework are 
discussed extensively in the contribution by Kotani \cite{R10}.  A 
reader is also referred to publications by present author {\it et al.} 
\cite{R5,R11,R12,R13} where the latter issue is addressed.

\section{$3d$ transition metal compounds}
\begin{figure} [H]
\centerline{\epsfxsize=7.5cm \epsfbox{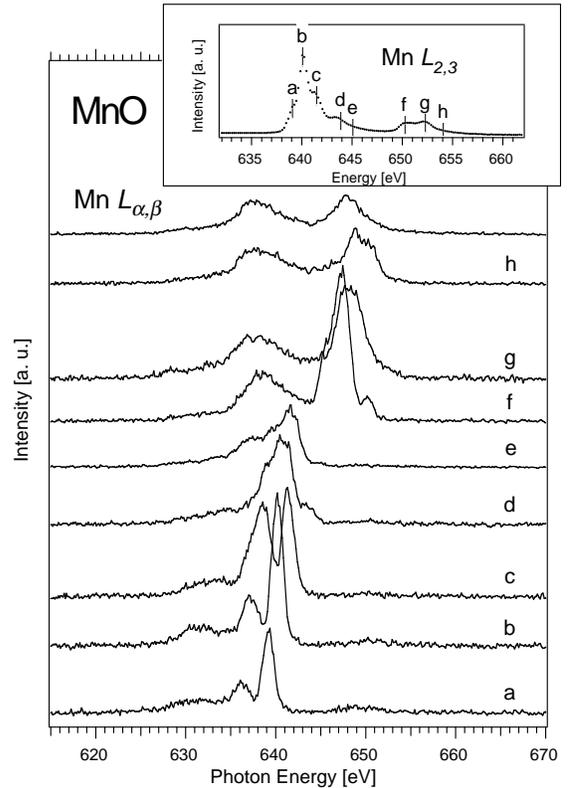}}
\caption{Mn $L_{2,3}$ x-ray fluorescence spectra of single-crystal MnO 
(100) recorded at different excitation energies near Mn 2$p$ 
thresholds (Ref.~\cite{R8}).  Excitation energies are indicated by 
ticks on the absorption spectrum shown in the inset.  The excitation 
for the uppermost spectrum was 716 eV.}
\label{fig3}
\end{figure}

\begin{figure} [H]
\centerline{\epsfxsize=8.0cm \epsfbox{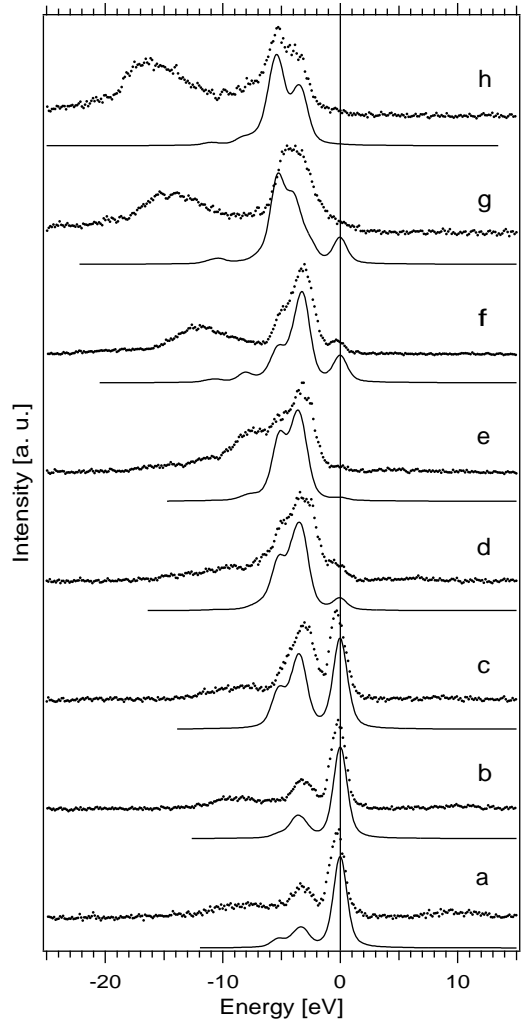}}
\caption{Resonant Mn $L_{2,3}$ x-ray fluorescence data (dots) of 
single-crystal MnO plotted as energy loss spectra together with 
results of model calculations (solid lines) for the Mn$^{2+}$ ion 
(Ref.~\cite{R8}).  Letters correspond to the same excitation energies 
as in Fig.~3.}
\label{fig4}
\end{figure}

The study of $d$-$d$ excitations is particularly important for 
elements from the middle of the 3$d$ row because a multiplet structure 
of the ground state configuration is fairly rich.  In the final state 
of the nonresonant x-ray emission and photoemission processes a system 
has one electron less compared to the ground state, as a consequence a 
multiplet structure observed in the x-ray emission and valence 
photoemission spectra is different from that for the ground state 
configuration.  Instead, optical absorption and EELS are usually used 
to study low-energy $d$-$d$ excitations.  These spectroscopies are 
however not element selective so that intra-atomic $d$-$d$ transitions 
often appear as weak structures on the slope of the main absorption 
edge.  Rich multiplets can be partly hidden under intense inter-band 
transitions. For instance, $d$-$d$ excitations of MnO with energies 
of around 5 eV and higher can hardly be observed in optical and EELS 
spectra \cite{R14} for that reason. Whereas, RIXS do not have this 
disadvantage as can be seen from the analysis of the data in 
Figs.~\ref{fig3} and \ref{fig4}. 

When the energy of incident photons is set at the Mn 2$p$ threshold an 
excited electron itself screens the core hole.  Due to dipole 
selection rules, there are different radiative transitions for the 
deexcitation process: back to the ground state or to low lying excited 
states so that the multiplet structure of the ground state 
configuration can be probed.  Mn $L_{2,3}$ ($3d,4s\rightarrow2p$ 
transitions) x-ray fluorescence spectra of single-crystal MnO (100), 
displayed in Fig.~\ref{fig3}, can be ordered by assigning the peaks to 
one of three categories: the recombination (elastic) peak, the 
resonating loss structures due to $d$-$d$ excitations and charge 
transfer transitions (i.~e.  RIXS structures), and the normal 
$L_{\alpha,\beta}$ x-ray emission lines which appear at constant 
energies of emitted photons.  The electronic recombination peak is at 
the same energy as the excitation energy, except for the possibility 
of phonon losses.

The relative intensity of the recombination peak decreases with 
increasing excitation energy.  This can be understood as a consequence 
of the spin ordering of the excited states.  The lowest $2p^53d^6$ 
intermediate states would have the highest possible spin (also 
sextets).  If these states decay, they are likely to end up again as a 
sextet, the $^6S$ ground state, contributing to the recombination 
line.  But at higher energies in the $L_3$ absorption multiplet one 
finds quartets and doublets (spin is not conserved because of the 
large

\noindent $2p$ spin-orbit interaction).  These higher-lying quartets 
are less likely to recombine into the sextet ground state, which 
explains the relatively weak recombination line.  This general 
ordering of the spin states in the intermediate state is not only true 
within the $L_3$ region, but also over the whole spectrum, as 
explained by Thole and Van der Laan \cite{R15}.  Within the $L_2$ 
region, one can see a clearly lower intensity of the elastic peak on 
the high-energy slope (Fig.~\ref{fig3}$h$) than on the first maximum 
(Fig.~\ref{fig3}$f$).  Another qualititative explanation for the 
trends in the intensity of the elastic peak is that if the absorption 
is relatively weak, the optical matrix element from the excited state 
back to the ground state is also small, and the core hole is more 
likely to decay to an excited final state.

Likewise the elastic peak, inelastic scattering structures reveal the 
strong dependence on varying excitation energies.  To make it easier 
to identify excitations in the final state, the Mn $L_{2,3}$ x-ray 
fluorescence data are displayed on the energy-loss scale in 
Fig.~\ref{fig4} along with results of model atomic-multiplet 
calculations for the Mn$^{2+}$ ion from Ref.~\cite{R8}.  The 
calculations were performed using equation~(2) where interference 
terms were neglected.  In other words, intensities were summed 
incoherently.  The atomic multiplet structures and matrix elements of 
dipole transitions between $2p^53d^6$ and $3d^5$ configurations were 
calculated in intermediate coupling in the spherical $O_3$ group using 
Cowan's programs \cite{R16}.  The radial parts of the 3$d$-3$d$ and 
2$p$-3$d$ Coulomb and exchange multipole interactions (so-called 
Slater integrals) were scaled down to 80\% of their Hartree-Fock 
values to account for intra-atomic configuration interaction and 
hybridization effects.

One can see in Fig.~\ref{fig4} that these calculations of 
$3d^5\rightarrow2p^53d^6\rightarrow3d^5$ transitions are very 
successful in reproducing the spectra excited on the $L_3$ absorption 
multiplet (spectra $a,b,c,d$).  Although the crystal field interaction 
was neglected, the atomic approach is adequate for the interpretation 
of experimental data due to an extra stabilization of the $3d^5$ 
high-spin configuration by the Hund's rule coupling.  Such an 
intra-atomic exchange stabilization results in a large energy 
separation between the ground high-spin $^6S$ and first excited 
low-spin $^4G$ states \cite{R17} compared to the crystal field 
splitting.  The first distinctly-resolved inelastic scattering 
structure in resonant Mn $L_{2,3}$ x-ray fluorescence spectra of MnO 
is observed at an energy loss of about 3 eV. According to the 
atomic-multiplet calculations, this structure corresponds to the 
transitions to $^4G$, $^4P$, and $^4D$-derived states.  When the 
crystal-field interaction is taken into account (see e.~g.  
Ref.~\cite{R18}), the $\sim$3~eV structure can be described to have 
contributions of $^4T_{1g}$, $^4T_{2g}$, $^4A_{1g}$, and $^4E_{g}$ 
symmetries.  Futher increase in the excitation energy results in a 
development of a shoulder at a loss energy of about 5 eV 
(Fig.~\ref{fig4}, spectra $c,d$) which is mainly composed by 
transitions to states with $^4A_{2g}$, $^4T_{1g}$, and $^4T_{2g}$ 
character.  Spectral weight in this energy-loss region becomes 
strongly enhanced for excitation energies set to the $L_3$ absorption 
multiplet.  In addition, non-zero intensity can be observed for 
$d$-$d$ excitations within $\sim$10 eV below the recombination peak 
despite some ovelap of inelastic scattering spectra with the 
$L_\alpha$ emission line at the fixed energy of 638 eV (increasing 
loss energy).  The latter line appears as a result of excitations into 
the $L_3$ continuum and Coster-Kronig decay from the $L_2$ hole 
states.  As a whole, the RIXS data obtained for MnO indicate that the 
RIXS technique offers an opportunity to study $d$-$d$ excitations in 
the extended energy range which are often not accessible with optical 
spectroscopy and EELS.

The natural extension of the RIXS spectroscopy to ease the symmetry 
identification for elementary excitations is polarization-dependent 
measurements.  While the reader is referred to papers by Duda {\it et 
al.} \cite{R19,R20} and Hague {\it et al.} \cite{R21}, which are in 
the same issue of the journal, for more extensive description of 
linear and circular magnetic dichroism studies, here Fe $L_{2,3}$ 
x-ray fluorescence data of single-crystal FeCO$_{3}$ from 
Ref.~\cite{R20} are used as an example.  We show the success of 
crystal-field multiplet calculations in reproducing of structures in 
resonant spectra due to the dipole nature of the spectroscopic 
process.  We provide the evidence of that, for the excitation close to 
Fe $2p$ thresholds, the experimental spectra of this compound are 
entirely dominated by the x-ray scattering contribution with the 
$L_\alpha$ normal emission intensity being significant only when the 
$L_3L_2M_{4,5}$ Coster-Kronig decay channel is open.

\begin{figure} [H]
\centerline{\epsfxsize=8.0cm \epsfbox{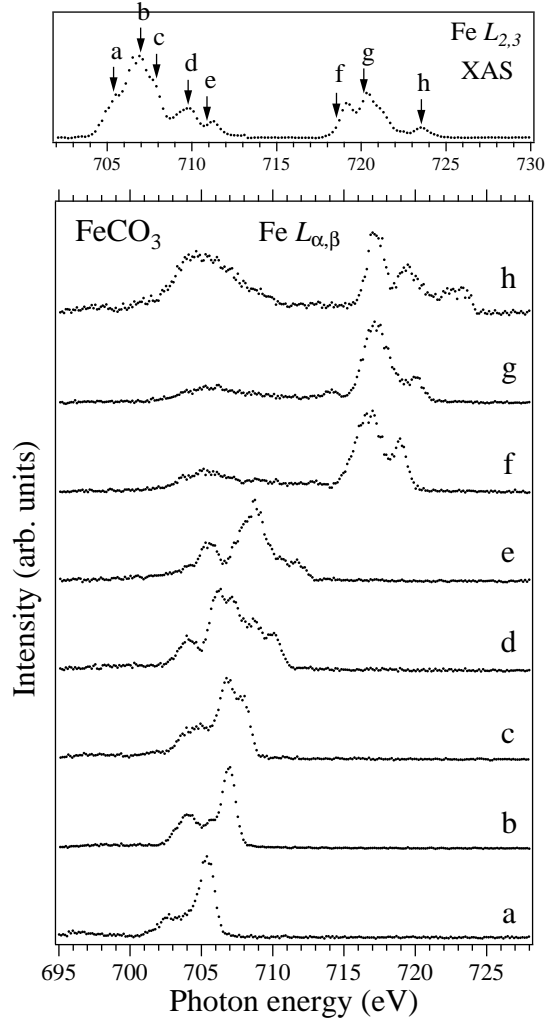}}
\caption{Fe $L_{2,3}$ x-ray fluorescence spectra of single-crystal 
FeCO$_3$ recorded at different excitation energies across Fe 2$p$ 
thresholds with the polarization vector ${\mathbf{E}}_{in}$ of 
incident photons parallel to the $\mathbf{c}$ axis of the crystal 
(adopted from Ref.~[20).  The spectra are normalized to the same 
height.  Excitation energies are indicated by arrows on the absorption 
spectrum shown in the top panel.}
\label{fig5}
\end{figure}

\begin{figure} [H]
\centerline{\epsfxsize=8.0cm \epsfbox{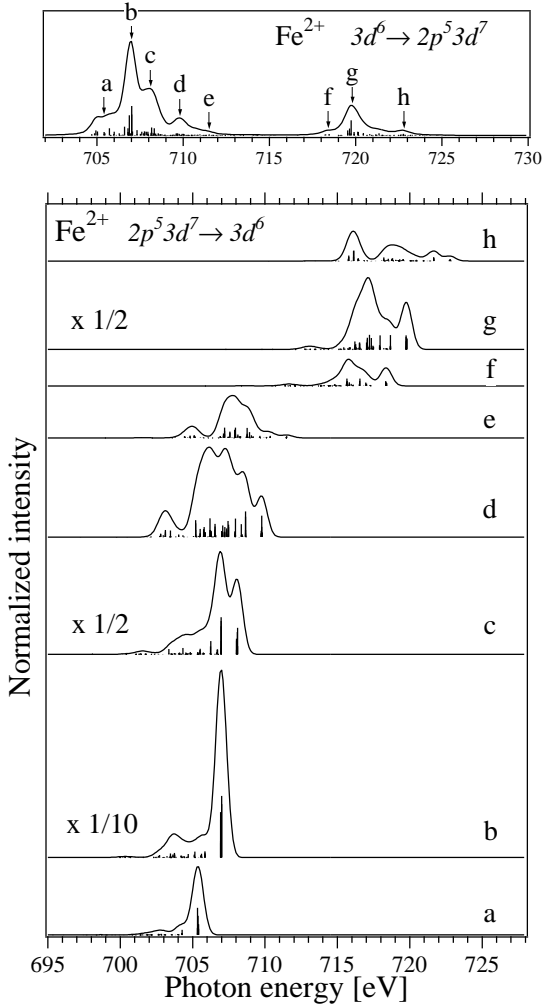}}
\caption{Results of crystal-field multiplet calculations of spectra 
displayed in Fig.~5. Spectral profiles are calculated for zero 
temperature.}
\label{fig6}
\end{figure}

Experimental data, obtained with the polarization vector 
${\mathbf{E}}_{in}$ of incident photons being parallel to the 
$\mathbf{c}$ axis of the FeCO$_{3}$ crystal, are displayed in 
Fig.~\ref{fig5}.  The corresponding calculated spectra for the 
Fe$^{2+}$ system are snown in Fig.~\ref{fig6}.  The calculations were 
performed using equation~(2) within the framework of crystal-field 
multiplet theory.  Slater integrals and matrix elements were obtained 
using Cowan's \cite{R16} and Butler's \cite{R22} codes, respectively, 
which were modified by Thole \cite{R23}.  A 20\% reduction was applied 
to Hartree-Fock values of Slater integrals.  Regarding a small 
trigonal distortion of FeCO$_{3}$, the calculations were done in the 
basis for $C_{3v}$ symmetry, although crystal-field parameters 
$D\sigma$ and $D\tau$ were set to very small values, for simplicity.  
The 10$Dq$ parameter was equal to 1.1 eV. The polarization of incident 
photons was taken to be along the trigonal axis with the 90$^\circ$ 
angle between directions of the incoming and outgoing radiation.  The 
lifetime $\Gamma$ of the intermediate state was set to 0.2 and 0.4 eV 
for $L_3$ and $L_2$, respectively.

We find that calculated RIXS spectra are very sensitive to the value 
of the 10$Dq$ parameter.  Distinct splittings observed in experimental 
spectra and reproduced in calculations with 10$Dq=1.1$~eV become 
obscure in calculated profiles at 10$Dq=1.0$~eV. The most sensitive 
spectrum is the one for the excitation $d$.  Its highest structure on 
the low photon energy side (at $\sim$706.1 eV in Fig.~\ref{fig6}) 
shows the extreme dependence on small variations of 10$Dq$, thus 
providing a good fingerprint of the crystal-field strength.  Indeed, 
the 10$Dq$ value derived in present calculations is consistent with 
estimations from other publications (see e.~g.  Ref.~\cite{R24}).

A very good agreement between calculated resonant x-ray scattering 
spectra of Fe$^{2+}$ and the resonant part of experimental Fe 
$L_{2,3}$ x-ray fluorescence data of FeCO$_{3}$ indicates a 
significantly low contribution to the spectra from charge-transfer 
excitations in the latter system, which were not taken into account in 
the calculations, as well as from normal emission.  In fact, a sizable 
contribution of normal emission in the $L_3$ region is observed only 
for excitation energies set to the $L_2$ edge as a result of the 
Coster-Kronig decay of the $L_2$ hole.  The ionic character of Fe 
chemical bonds in FeCO$_{3}$ enables crystal-field theory to describe 
RIXS data in detail and as a conse-

\noindent quence to provide knowledge about the ground state and 
low-lying excited states.

For highly covalent compounds, it is however necessary to take into 
account charge-transfer excitations and configurational mixing in the 
ground and intermediate states of the spectroscopic process in 
analysis of experimental data.  The configuration interaction modifies 
(sometimes significantly) the spacing between energy levels resulting 
from electrostatic, crystal field, spin-orbit, exchange interactions, 
etc.  The character of states, expressed as a linear combination of 
wave functions, may change significantly as well.  Charge-transfer 
effects can produce intense structures (charge-transfer satellites) in 
RIXS spectra, the energy-loss of which is related to physical 
quantities being crucial for understanding of ground state properties.  
In particular, the charge-transfer energy 
$\Delta=\varepsilon_d-\varepsilon_p$, required for a transfer of an 
electron from a ligand site to a metal site, and hybridization 
strength $V$ between metal $3d$ states and ligand states are employed 
in the description of coupling between the ground state configuration 
$3d^n$ and excited $3d^{n+1}\underline{L}$ ($\underline{L}$ denotes a 
hole in the valence band), thus yielding the 
$\sqrt{\Delta^2+4V_{\mathrm {eff}}^2}$ separation between bonding and 
anti-bonding states.  $U$ for $3d$ electrons is also included in the 
description of coupling between $3d^{n+1}\underline{L}$ and 
$3d^{n+2}\underline{L}^2$ configuration to define the $\Delta+U$ 
separation between gravity centers of these configurations in the 
limit of $V\rightarrow0$.  Charge-transfer satellites in RIXS spectra 
corresponding to transitions to states of predominantly 
$3d^{n+1}\underline{L}$ or $3d^{n+2}\underline{L}^2$ character can be 
strongly enhanced by setting the excitation energy to their 
$2p^53d^{n+2}\underline{L}$ or $2p^53d^{n+3}\underline{L}^2$ 
conterparts in metal $2p$ absorption spectra.  Due to this resonant 
behavior, energy positions of satellites can be determined more 
accurately, thus resulting in more accurate estimations of related 
physical quantities.

\begin{figure} [H]
\centerline{\epsfxsize=8.0cm \epsfbox{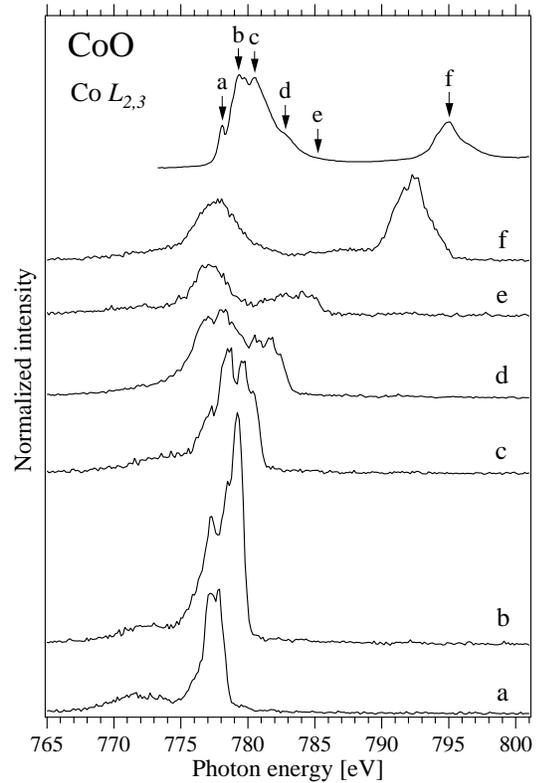}}
\caption{Co $L_{2,3}$ x-ray fluorescence spectra of single-crystal CoO 
(100) recorded at different excitation energies near Co 2$p$ 
thresholds (Ref.~\cite{USX}).  Excitation energies are indicated by 
arrows on the absorption spectrum shown on top.}
\label{fig7}
\end{figure}

\begin{figure} [H]
\centerline{\epsfxsize=8.5cm \epsfbox{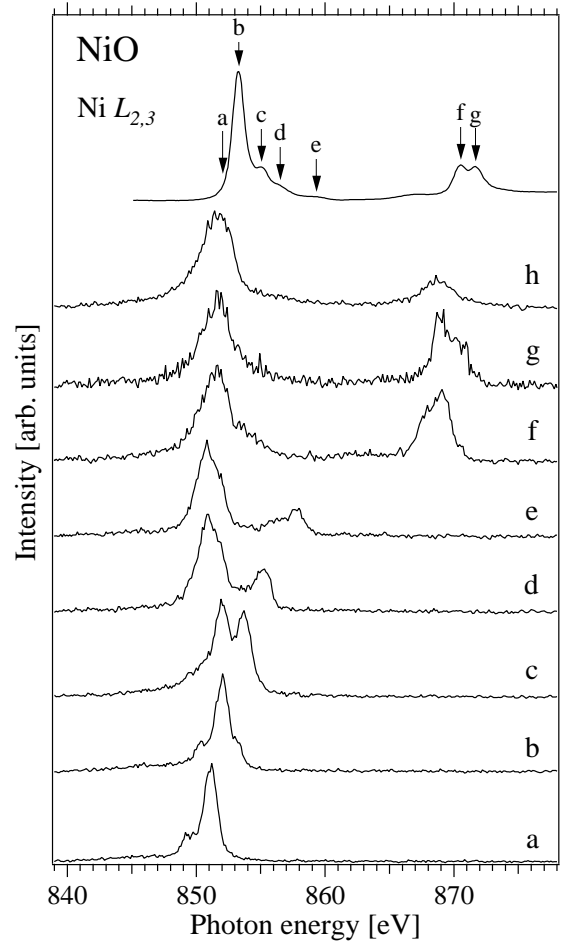}}
\caption{Ni $L_{2,3}$ x-ray fluorescence spectra of single-crystal NiO 
(100) recorded at different excitation energies near Ni 2$p$ 
thresholds (Ref.~\cite{USX}).  Excitation energies are indicated by 
arrows on the absorption spectrum shown on top.  The excitation for 
spectrum $h$ was 916.5 eV.}
\label{fig8}
\end{figure}

\begin{figure}
\centerline{\epsfxsize=8.5cm \epsfbox{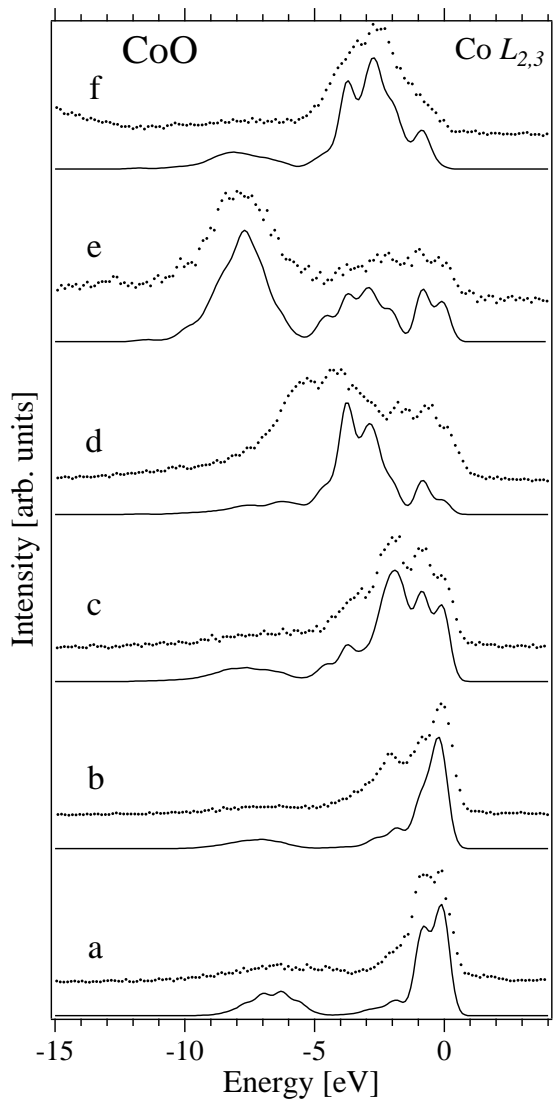}}
\caption{Experimental x-ray scattering spectra of single-crystal CoO 
(100) at Co $2p$ thresholds (dots) together with the results of 
Anderson impurity model calculations for the Co$^{2+}$ system in 
octahedral symmetry.  Letters correspond to the same energies of 
incident photons as in Fig.~7.}
\label{fig9}
\end{figure}

\begin{figure}
\centerline{\epsfxsize=8.5cm \epsfbox{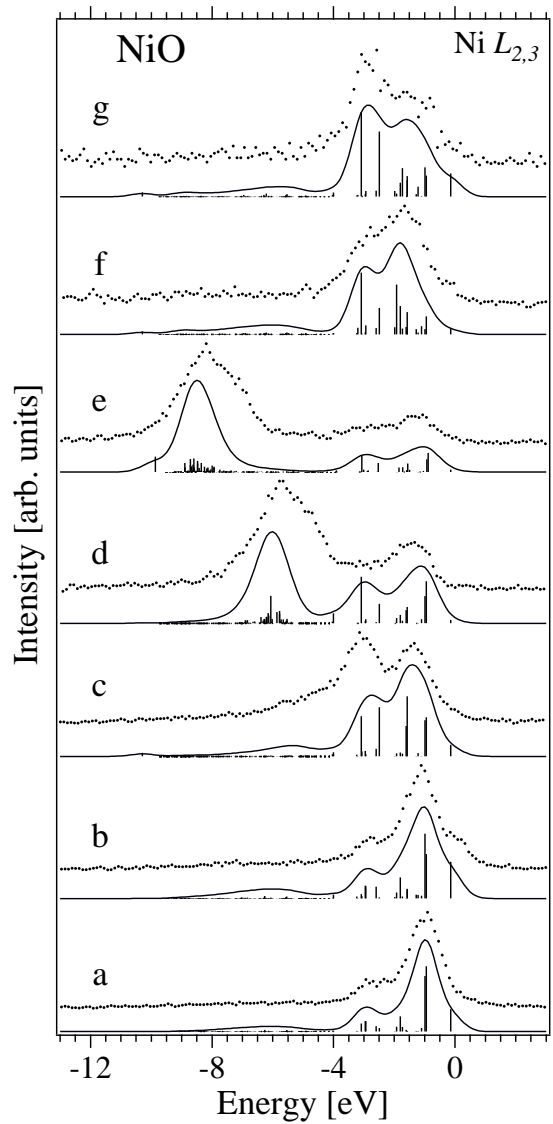}}
\caption{Experimental x-ray scattering spectra of single-crystal NiO 
(100) at Ni $2p$ thresholds (dots) together with the results of 
Anderson impurity model calculations for the Ni$^{2+}$ system in 
octahedral symmetry.  Letters correspond to the same energies of 
incident photons as in Fig.~8.}
\label{fig10}
\end{figure}

In Figs.~\ref{fig7} and \ref{fig8}, resonant Co and Ni $L_{2,3}$ x-ray 
fluorescence spectra of CoO and NiO, respectively, are shown on the 
photon energy scale, covering the corresponding absorption edges.  
Besides the structures clearly dispersing on this scale with varying 
excitation energies there are lines which barely show any energy 
shifts.  While the dispersing structures can be tentatively attributed 
to $d$-$d$ excitations the assingment of other lines to normal 
emission is not that obvious for different spectra.  In particular, 
for CoO, setting the excitation energy to the high-energy tail of the 
$L_3$ absorption line gives rise to an appearance of an intense line 
at about 777 eV in spectrum $e$ in Fig.~\ref{fig7}.  The energy 
position of this line is somewhat different from that of $L_\alpha$ in 
spectrum $f$, thus indicating that it can not entirely consist of the 
normal emission contribution.  The effect is more pronounced for NiO. 
Low photon energy lines in spectra $d$ and $e$ in Fig.~\ref{fig8} 
reveal a distinct low energy shift relative to $L_\alpha$ recorded at 
an excitation energy set far above $2p_{3/2}$ and $2p_{1/2}$ 
thresholds.  The results of Anderson impurity model calculations (see 
Figs.~\ref{fig9} and \ref{fig10}) suggest that these lines mainly 
originate from O~$2p\rightarrow$~metal~$3d$ charge-transfer 
excitations.

Calculations were done using formula~(2).  Matrix elements were 
obtained by applying a chain of programs which, in addition to Cowan's 
and Butler's codes, comprises a charge-transfer program written by 
Thole and Ogasawara.  The $3d^n$ and $3d^{n+1}\underline{L}$ 
configurations for the ground and final states of the spectroscopic 
process and the $2p^53d^{n+1}$ and $2p^53d^{n+2}\underline{L}$ 
configurations for the intermediate state were included in these 
calculations ($n$ is equal to 7 and 8 for CoO and NiO, respectively).  
The $3d^{n+2}\underline{L}^2$ configuration was not taken into account 
in the description of the ground state because its contribution was 
estimated to be only $\sim$1\% in these oxides \cite{R25,R26}.  
Spectra, displayed in Figs.~\ref{fig9} and \ref{fig10}, were 
calculated for the incident light linearly-polarized along the 
$z$-axis and 90$^\circ$ detection geometry with the parameter values 
summarized in Table~1.  The intermediate state lifetime $\Gamma$ was 
set to 0.5 and 0.7 eV at $L_3$ and $L_2$, respectively, for CoO and to 
0.6 and 0.8 eV for NiO. The hybridization strength $V$ was taken to be 
weaker in the intermediate state than that in the ground and final 
states due to its significant configurational dependence, as {\it a 
priori} expected \cite{R3}. 

One can see in Fig.~\ref{fig9} that the discussed above line in 
spectrum $e$ of CoO (here it is at $\sim$8 eV on the energy loss 
scale) is reproduced by calculations as a resonance of transitions to 
the states of essentially $3d^{8}\underline{L}$ character.  Similar 
situation occurs for NiO. The intense lines in spectra $d$ and $e$ at 
the energy loss of $\sim$5.7 and $\sim$8.2 eV, respectively, were 
found to have mainly the $3d^{9}\underline{L}$ origin (see 
Fig.~\ref{fig9}) and represent the enhanced inelastic scattering 
structure due to charge-transfer excitations.  This structure does not 
follow an increasing excitation energy and is observed in the spectra 
rather at the constant photon energy than at the constant energy loss.  
Such behavior is not predicted by the simplistic considerations of the 
scattering process, described in Introduction. 

\noindent Nevertheless it is not 
surprising.  A possibility of the non-linear dispersion of inelastic 
scattering structures, corresponding to charge-transfer excitations, 
was discussed earlier by Tanaka~{\it et al.}~\cite{R27}.  It was shown 
that for the excitation energy set to the region of charge-transfer 
satellites in the absorption spectrum (in this case, intermediate 
states of mainly $2p^53d^{n+2}\underline{L}$ character), the 
corresponding charge-transfer satellite spectrum for inelastic 
scattering can be decoupled for $\Omega$ and $\omega$ in the limit of 
$\Gamma\rightarrow0$.  No correlation between incident and emitted 
photon energies can be observed for any changes of $\Omega$ within a 
range of $\Delta\Omega<W$ so that the charge-transfer structure 
behaves as a normal-emission-like line.

Present calculations also reveal the sensitivity of RIXS profiles to 
the non-zero exchange field giving rise to the spectral weight 
transfer \cite{R28,R29}.  Inter-atomic exchange interactions which 
correspond to very strong effective magnetic fields and act only on 
the valence electron spins were taken into account using a mean-field 
theory treatment.  In principle, the exchange field strength can be 
different in the ground and intermediate states of the spectroscopic 
process.  For example, for a charge-transfer gap material like NiO, 
the (super)exchange parameter, which is defined as 
$J\approx2V_{e_g}^4/\Delta_{\mathrm {eff}}^3$ \cite{R30}, is expected 
to be larger in the presence of a core hole due to the smaller 
effective $\Delta$.  In the calculations, the exchange fields in the 
ground and intermediate states were set to be the same, for 
simplicity.  The value of the exchange field used in spectral 
simulations for CoO seems to be too high since the N\'eel temperature 
in this oxide is lower than that in NiO. However, it is very difficult 
to reproduce experimental spectra of CoO without turning on the 
exchange field even if the Boltzmann distribution in the population of 
low-lying excited states at room temperature is taken into account in 
the calculations.  One possible reason for applying of an excessive 
exchange field may be an underestimation of spin-spin correlation 
effects on the spectral shape \cite{R31} in the molecular-field 
approximation.  Another reason could be a possible existance of Co 
vacancies in the bulk of the single-crystal so that induced electronic 
holes, having oxygen character, would be antiferromagnetically coupled 
with Co$^{2+}$.  The exchange interaction between these holes and 
nearest-neighbor Co ions is expected to be much stronger than the 
Co--Co (super)exchange interaction \cite{R32}, thus producing larger 
spectral effects and requiring larger effective exchange fields for 
the spectral simulations.

As a whole, results of Anderson impurity model calculations are in 
good agreement with experimental data except for spectrum $d$ of CoO. 
Although, all the structures are reproduced, their calculated relative 
intensities are not entirely correct.  In particular, the structure at 
the energy loss of $\sim$5.6 eV, which has a charge-transfer origin, 
is clearly lacking an intensity in the calculated spectrum.  This 
suggests that the difference $Q-U$ and/or bandwidth $W$, as parameters 
to large extent defining the charge-transfer satellite weight in the 
absorption spectrum at the corresponding excitation energy, may be 
somewhat larger than those used in calculations.

Despite the relative success of Anderson impurity model calculations 
of RIXS data in CoO and NiO, it is not really clear to what extent 
this approach may be valid for even more covalent Co 

\begin{figure}
\centerline{\epsfxsize=8.0cm \epsfbox{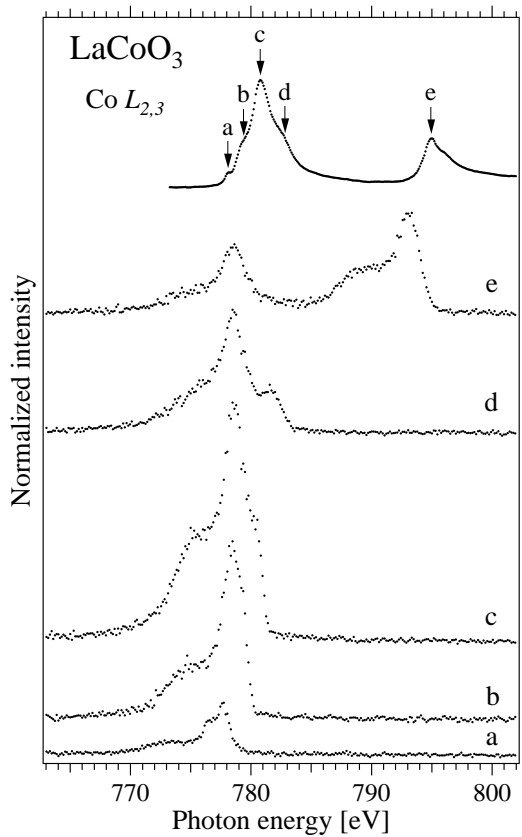}}
\caption{Co $L_{2,3}$ x-ray fluorescence spectra of LaCoO$_3$ recorded 
at different excitation energies near Co 2$p$ thresholds 
(Ref.~\cite{USX}).  Excitation energies are indicated by arrows on the 
absorption spectrum shown on top.}
\label{fig11}
\end{figure}

\begin{figure}[H]
\centerline{\epsfxsize=8.0cm \epsfbox{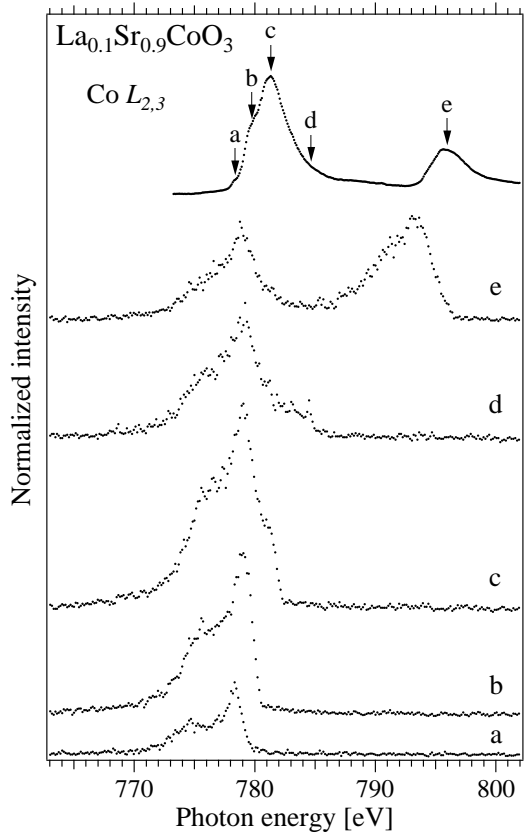}}
\caption{Co $L_{2,3}$ x-ray fluorescence spectra of 
La$_{0.1}$Sr$_{0.9}$CoO$_3$ recorded at different excitation energies 
near Co 2$p$ thresholds (Ref.~\cite{USX}).  Excitation energies are 
indicated by arrows on the absorption spectrum shown on top.}
\label{fig12}
\end{figure}

\begin{figure} [H]
\centerline{\epsfxsize=8.0cm \epsfbox{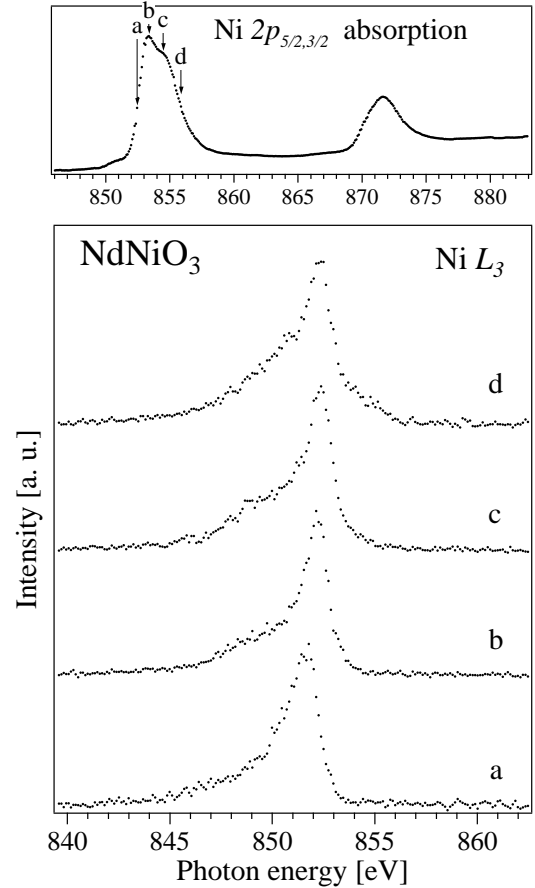}}
\caption{Ni $L_{3}$ x-ray fluorescence spectra of NdNiO$_3$ recorded 
at different excitation energies near the Ni 2$p_{5/2}$ threshold 
(Ref.~\cite{USX}).  Excitation energies are indicated by arrows on the 
absorption spectrum shown in the top panel.}
\label{fig13}
\end{figure}

\begin{table} [H]
\centering
\caption{Values for parameters used in Anderson impurity model 
calculatins of resonant x-ray scattering spectra at metal $2p$ 
thresholds in CoO and NiO, where $\kappa$ is a scaling coefficient for 
Slater integrals, $\Delta$ is defined as an energy difference between 
gravity centers of $3d^{n+1}\underline{L}$ and $3d^n$ configurations, 
$V_{e_{g}}$ represents hopping for $e_{g}$ orbitals ($V_{t_{2g}}$ 
taken as half of the $V_{e_{g}}$ value), $W$ is the width of the O 
$2p$ band which shape is approximated by a circle, $N$ is the number 
of levels in the valence band, $Q$ is a core-hole potential.  The 
inter-atomic-exchange field is applied along the $z$-axis.  All the 
values, except for those for $\kappa$ and $N$, are in units of eV.}
\bigskip
\begin{tabular}{lcr}
\hline
Parameters&CoO&NiO\\
\hline
$\kappa$&0.8&0.8\\
$\Delta$&4.0&3.5\\
$V_{e_{g}}$, ground state&2.2&2.2\\
$V_{e_{g}}$, intermediate state&1.8&1.8\\
$W$&4.0&5.0\\
$N$&8&10\\
$Q-U$&0.0&1.0\\
$10Dq$&0.5&0.5\\
Exchange field&0.3&0.15\\
\hline
\\
\end{tabular}
\label{table1}
\end{table}

\noindent and Ni systems, 
where $3d$ states have higher degree of delocalization.  In fact, for 
highly covalent compounds, such as La$_{1-x}$Sr$_x$CoO$_3$ and 
NdNiO$_3$, only small changes in the shape of Co and Ni $L_{2,3}$ 
x-ray fluorescence spectra with varying excitation energies across the 
corresponding $2p$ absorption edges are observed (see 
Figs.~\ref{fig11},~\ref{fig12}, and ~\ref{fig13}).  Although 
relatively high resolution was used to measure these data (total 
resolution was set to about 0.7 eV for both Co and Ni compounds), the 
spectra do not show many structures and look quite broad.  The energy 
position of the main peak has almost no dependence on the excitation 
energy tuned above or at the $2p$ threshold.  However, at least for 
La$_{0.1}$Sr$_{0.9}$CoO$_3$ and NdNiO$_3$, the interpretation of these 
spectra as mostly consisting of a normal emission contribution, which 
would be a projection of the partial density of states, can be 
challenged by their Anderson impurity model description as spectra of 
compounds with negative $\Delta$.  In the latter case, the ground and 
low-lying excited states will mainly have the $3d^{n+1}\underline{L}$ 
character so that main spectral structures may behave in a similar 
manner as charge-transfer satellites in the RIXS spectra of CoO and 
NiO, i.~e.  reveal no shifts on the emitted photon energy scale for 
some range of incident photon energies.  Detailed calculations of 
spectral shapes using both band theory and quasi-atomic approach are 
needed to clarify the role of partial $3d$ localization and $d$-$d$ 
correlation in these systems.

\section{Lanthanide compounds}
Since the nature of the ground state defines various physical 
properties of a system in question, knowing the ground state character 
is a key issue, especially in materials science.  It turns out that 
for rare-earth compounds the RIXS technique can provide information 
about $J$-mixing in the ground state through studies of intra-atomic 
$f$-$f$ excitations.  

When the symmetry in a solid is not spherical, the angular momentum of 
$f$ electrons is not conserved and therefore $j$ is not a good quantum 

\noindent number.  However, the crystal field is an order of magnitude 
smaller than the spin-orbit splitting and can be treated as a 
perturbation.  In the case of weak hybridization effects, the 
interlevel coupling and consequently $J$-mixing in the ground state of 
the system are often disregarded in the interpretation of experimental 
data by applying a pure atomic approximation (mainly for high-energy 
spectroscopies) or by using a first order crystal-field theory where 
the crystal field interaction is assumed to act only within the 
separate $J$ manifolds.  This is partly due to complications in 
extracting information about the ground state $J$-mixing directly from 
the data.  For example, the estimation of the $J$-mixing degree in 
high-order crystal-field theory by adjusting the crystal-field 
parameters from the fit of optical absorption or low-energy 
electron-energy-loss spectra \cite{A1,A2} may result in a large 
uncertainty originating from difficulties calculating the intensities 
of dipole-forbidden transitions.  In turn, the possible influence of 
the weak metal-ligand hybridization is difficult to analyze 
quantitatively in the absence of so-called charge-transfer satellites 
in high-energy spectroscopic data.

In this situation, the use of alternate spectroscopic means to obtain 
ground-state $J$-mixing information is essential.  Recently, Finazzi 
{\it et al.} \cite{A3} have shown that this mixing can be studied by 
taking advantage of dichroic properties of rare-earth $3d$ x-ray 
absorption.  However, the method is limited to magnetically ordered 
systems.  Here, we discuss a potential of resonant x-ray scattering 
spectroscopy in studying of the ground-state $J$-mixing when applied 
to compounds without distinct long-range magnetic order and 
significant metal-ligand hybridization.

Similar to optical absorption and EELS with respect to probing the 
optical scale excitations, RIXS at the same time provides an 
additional level of the transition selectivity due to the element 
specificity and dipole selection rules.  In contrast to systems with 
the strong metal-ligand hybridization where the charge-transfer 
process leads to an appearance of intense lines in RIXS spectra as a 
result of inter-ionic excitations, $J$-mixing in systems with weak 
hybridization effects is expected to manifest itself in an intensity 
gain of some intra-ionic ($f$-$f$) transitions which are disallowed 
for the pure Hund-rule ground state.  In other words, transitions with 
$\Delta{J}$ other than 0, $\pm$1, and $\pm$2 are probed in the 
resonant excitation-deexcitation process.  Although $J$ is not a good 
quantum number in the $J$-mixing case, we use this terminology for 
simplicity.

\begin{figure} [H]
\centerline{\epsfxsize=7.0cm \epsfbox{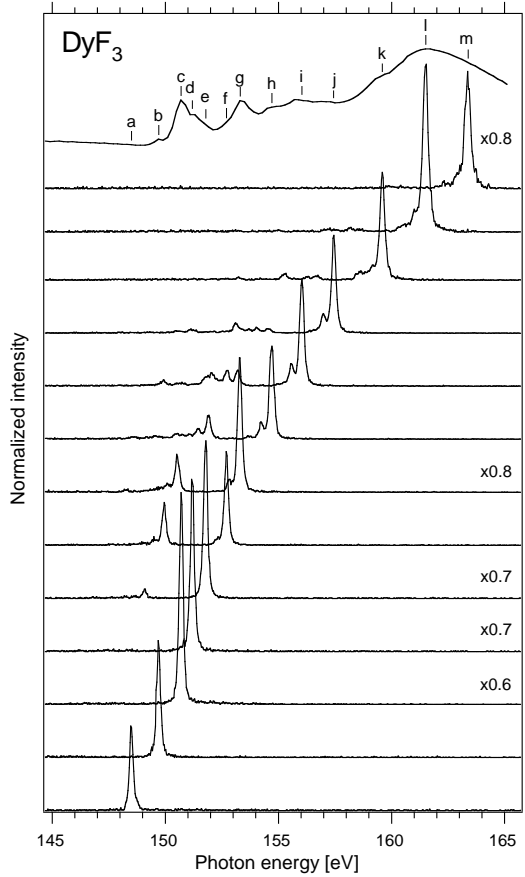}}
\caption{The total electron yield spectrum at the Dy $4d$ edge and 
resonant x-ray scattering spectra of DyF$_3$ normalized to the 
incident photon flux (Ref.~\cite{ALS98}).  The letters correspond to 
the excitation energies indicated in the absorption spectrum.}
\label{fig14}
\end{figure}

\begin{figure} [H]
\centerline{\epsfxsize=7.0cm \epsfbox{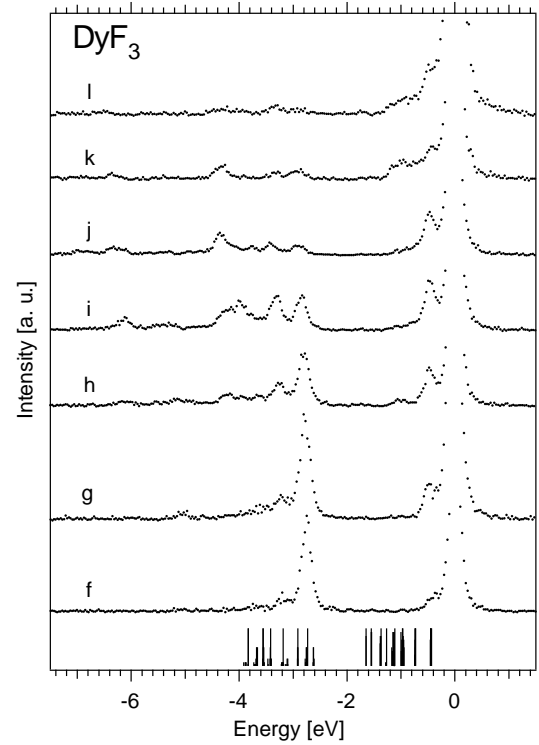}}
\caption{Enlarged inelastic x-ray scattering part of the resonant Dy 
4$f\rightarrow4d$ spectra from DyF$_3$ together with optical 
absorption transitions of Dy$^{3+}$ in LaF$_3$ at 4.2 K 
(Ref.~\cite{A7}).  For optical transitions, the most intense lines in 
each group of Stark levels belonging to each $J$ are all normalized to 
the same value.  The eight clearly separated manifolds with energies 
under 2 eV are assigned to $^6H_{13/2}$, $^6H_{11/2}$, ($^6H_{9/2}$, 
$^6F_{11/2}$), ($^6H_{7/2}$, $^6F_{9/2}$), $^6H_{5/2}$, $^6F_{7/2}$, 
$^6F_{5/2}$, and $^6F_{3/2}$ (in order of the increasing 
transition energy), respectively.}
\label{fig15}
\end{figure}

A discussion about the RIXS potential to probe the ground-state 
$J$-mixing is based on analysis of data obtained at the Dy $4d$ 
threshold of DyF$_{3}$ at room temperature.  Measurements at the $4d$ 
threshold of rare-earths provide naturally higher resolution than 
those at the $3d$ threshold, thus allowing one to study elementary 
excitations in greater detail (see e.~g.  Ref.~\cite{A4}).  
Experimental spectra of DyF$_{3}$ are displayed in Figs.~\ref{fig14} 
and ~\ref{fig15} on both photon-energy and energy-loss scales.  Two 
distinct groups of pronounced inelastic-scattering peaks are observed 
in these spectra.  The first group is distinguished by small energy 
losses on the tail of the elastic line, whereas the second is 
characterized by energy losses more than 2.5 eV. When the excitation 
energy approaches the main broad maximum of the Dy $4d$ absorption 
edge, the first group still possesses significant intensity while the 
structures of the second group become relatively faint.  Regarding the 
energy scale on which the spectral variations occur, the observed 
transitions can be attributed to intra-ionic $f$-$f$ excitations.  The 
energy gap between two groups of inelastic x-ray scattering structures 
reflects the separation between sextets and quartets of trivalent Dy 
\cite{R17,A5} which can be reached due to the excitation-deexcitation 
process.

Indeed, the overall spectral shapes and their behavior with varying 
excitation energies are reasonably well reproduced by atomic multiplet 
calculations for the Dy$^{3+}$ ion \cite{A6}.  The results of 
calculations show that the dominant elastic peak in all of the RIXS 
spectra of DyF$_{3}$ is to large extent a consequence of strong 
interference effects in the intermediate state of the coherent 
second-order optical process.  The states constituting the main 4$d$ 
absorption edge have a lifetime broadening of about 2 eV largely 
because of the $4d-4f4f$ Coster-Kronig decay.  However, a close 
inspection of experimental RIXS spectra shows that there are some 
spectral structures which are not revealed in calculations within the 
pure atomic approximation.  Thus, the feature with the energy loss of 
about 1.17 eV is observed in spectra $h$, $k$, and $l$, presented in 
detail by Fig.~\ref{fig15}.  While atomic multiplet theory predicts 
the non-zero intensities for resonant inelastic x-ray scattering 
transitions to the $^6H_{13/2}$, $^6H_{11/2}$, and $^6F_{11/2}$ 
sextets of the $4f^9$ configuration (the Hund rule's ground state is 
$^6H_{15/2}$), the energy of the extra-feature in experimental spectra 
$h$, $k$, and $l$ 

\noindent (Fig.~\ref{fig15}) is close to those of $^6F_{9/2}$ 
and $^6H_{7/2}$ manifolds of Dy$^{3+}$ in LaF$_3$ \cite{A7}.  This is 
an indication of $J$-mixing and the presence of $J=13/2$ and $J=11/2$ 
components in the ground state of DyF$_{3}$.  Indications of other 
extra-structures missing from atomic multiplet calculations can be 
seen in the energy loss range between 1.2 and 2.0 eV, as in spectra 
$l$ and $m$.

To simulate the effect of $J$-mixing, we calculated transition 
intensities within the pure atomic approximation by choosing 
$^6H_{13/2}$ and $^6H_{11/2}$ as 

\noindent initial states for the scattering 
process.  The excitation energy was taken to be the same as that for 
experimental spectrum $l$ (since the intensity of extra-structures is 
higher for excitation energies at the main absorption edge) and the 
core-hole lifetime broadening was set to 2.0 eV. According to these 
calculations, the spectral weight at an energy loss of about 1.17 eV 
(transitions to $^6F_{9/2}$ and $^6H_{7/2}$) should constitute 56 \% 
and 217 \% of that at 1.0 eV (transitions to $^6F_{11/2}$) for 
$^6H_{13/2}$ and $^6H_{11/2}$ as initial states, respectively.  
Although, accounting for weak but finite Dy $4f$-F $2p$ hybridization 
may give rise to some changes in calculated relative intensities in 
both cases, it is clear that an admixture of the $J=11/2$ component in 
the ground state results in a stronger effect than that for the 
$J=13/2$ component.  Since the degree of $J$-mixing in DyF$_{3}$ is 
relatively low, the $J=11/2$ contribution to the ground states is 
expected to be comparable with the $J=13/2$ contribution in order to 
explain the noticeable weight of ``forbidden'' transitions in the 
inelastic scattering spectra in Fig.~\ref{fig15}.  This is not 
unusual.  For example, the $J=11/2$ component has been found to be 
comparable to the $J=13/2$ component in the ground state of 
Dy$^{3+}$-doped yttrium scandium gallium garnet \cite{A2} as a result 
of the crystal-field interaction.

\begin{figure} [H]
\centerline{\epsfxsize=7.0cm \epsfbox{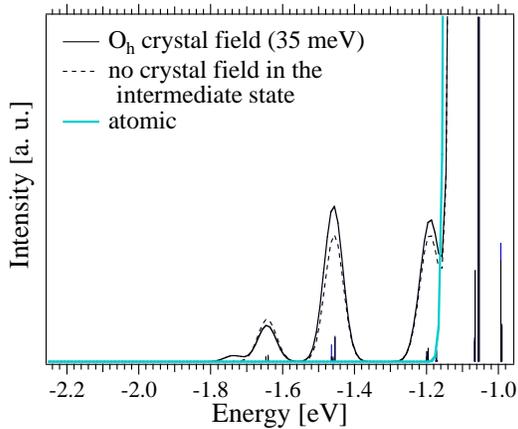}}
\caption{A 1.0-2.2 eV energy-loss region of inelastic x-ray scattering 
spectra of the Dy$^{3+}$ system calculated using a pure atomic 
approach (intermediate coupling) and crystal field multiplet theory 
($O_{h}$ symmetry).  In the calculations Slater integrals 
$F^k(5f,5f)$, $F^k(5d,5f)$, and $G^k(5d,5f)$ were scaled down to 80\%, 
75\%, and 66\%, respectively, from the Hartree-Fock values.  An 
excitation energy was set to the main absorption maximum (spectrum $l$ 
in Fig.~14) and $\Gamma$ was taken to be 2.0 eV.}
\label{fig16}
\end{figure}

To estimate the effect of this interaction on the shape of RIXS 
spectra, we also performed model crystal-field multiplet calculations 
for the Dy$^{3+}$ ion in the crystal field of $O_{h}$ symmetry with 
the strength of 35 meV. Fig.~\ref{fig16} shows the 1.0-2.2 eV 
energy-loss region of the spectra calculated using a pure atomic 
approximation and crystal field multiplet theory.  It is clear that 
switching on the crystal field gives rise to additional 
transitions.  The calculated intensities are low to fully account for 
the observed spectral weight in experimental data at the corresponding 
energy loss.  This suggests that inter-atomic coupling is also 
important for the description of the inelastic-scattering profile in 
the energy loss range between 1.0 and 2.0 eV and that the appearance 
of additional structures is rather a combined effect of the 
crystal-field interaction and Dy $4f$-F $2p$ hybridization.  However, 
the calculations which would take into account both the crystal field 
and F $2p\rightarrow$Dy $4f$ charge transfer excitations are 
complicated by a huge number of multiplets and require large 
computational resources.  At present, they are out of the scope of the 
paper.

The existance of $J$-mixing in the intermediate state raises a 
question about how strongly the inelastic-scattering intensity at 
energy losses between 1.0 eV and 2.0 eV is related to $J$-mixing in 
the ground state of DyF$_{3}$.  To estimate this, crystal-field 
multiplet calculations with the crystal-field interaction switched off 
in the intermediate state were performed.  Thus, any $J$-mixing in the 
core-excited state was disallowed.  A comparison of the results of 
calculations with and without crystal-field interaction in the 
intermediate state (Fig.~\ref{fig16}) shows no significant changes in 
the inelastic-scattering intensities of ``forbidden'' structures on 
switching off $J$-mixing in the core-excited state.  One of the main 
reasons for that is a large core-hole lifetime broadening.  As a 
whole, the calculations indicate that the spectral weight in the 
energy loss region between 1.0 and 2.0 eV is largely determined by 
$J$-mixing in the ground state of DyF$_{3}$.

\section{Actinide compounds}
The usual perspective on the actinides is that their $5f$ electrons 
are more localized than $3d$ electrons of transition elements but less 
localized than $4f$ electrons of lanthanides.  Charge-transfer effects 
are expected to be significant in actinide compounds as a result of 
metal 5$f-$ligand 2$p$ hybridization.  Indeed, the analysis of our 
data obtained at U $3d_{5/2}$ threshold shows that the ligand 
2$p\rightarrow$U 5$f$ charge-transfer plays an important role in 
uranium compounds, such as UO$_2$, UO$_2$(NO$_3$)$_2\cdot$6H$_2$O, and 
even in UF$_4$.

The most extensive studies of the electronic structure of these 
aforementioned compounds have been carried out for UO$_2$ 
(Refs.\cite{B1,B2}).  Molecular-orbital calculations by several 
research groups \cite{B3,B4,B5,B6} gave values for the 5$f$ occupancy, 
which range from 2.3 to 2.9 electrons, while this occupancy was 
estimated at about 2.3 electrons from the analysis of x-ray absorption 
and photoemission data within an Anderson impurity model \cite{B7,B8}.  
These results indicate significant degree of covalency for U--O 
chemical bonds in UO$_2$.  For UF$_4$, a $5f$ contribution of 
$\sim$0.3 electrons to the bonding orbitals was also predicted from 
relativistic Dirac-Slater local-density calculations \cite{B9}.  For 
compounds containing U$^{6+}$, the degree of covalency for 
metal$-$ligand bonds is expected to be even higher than that for 
U$^{4+}$ systems.  For example, molecular-orbital calculations yielded 
the 5$f$ occupancy of $\sim$2.6 electrons for the uranyl ion 
UO$_2^{2+}$ (Refs.  \cite{B10,B11}).  Although, the values for the 
5$f$ occupancy obtained from molecular-orbital calculations seem to be 
overestimated \cite{B12} one can not rule out the importance of the U 
5$f$-ligand 2$p$ hybridization even in a compound with ``ionic'' bonds 
such as UF$_4$.

\begin{figure}
\centerline{\epsfxsize=8.0cm \epsfbox{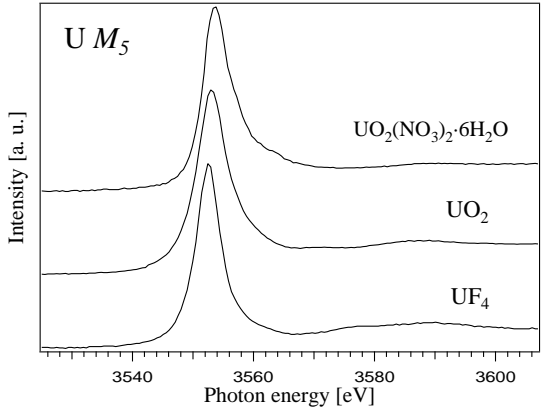}}
\caption{Total electron yield spectra of UF$_{4}$, UO$_{2}$, and
UO$_2$(NO$_3$)$_2\cdot$6H$_2$O near the U $M_{5}$ absorption edge.}
\label{fig17}
\end{figure}

As discussed above, one of the consequences of high covalency and 
hybridization in the ground state is an appearance of charge-transfer 
satellites in the high-energy spectroscopic data.  For actinides, the 
3$d$ core-hole lifetime broadening is quite large, thus reducing the 
efficiency of the x-ray absorption technique.  As a result, the U 
3$d_{5/2}$ x-ray absorption spectra of UF$_4$, UO$_2$, and 
UO$_2$(NO$_3$)$_2\cdot$6H$_2$O, displayed in Fig.~\ref{fig17}, do not 
exhibit many sharp features.  In particular, spectra of UF$_4$ and 
UO$_2$ appear as a single line with some asymmetry on the high-energy 
side (weak and broad structures in the continuum at about 3576 and 
3590 eV for UF$_4$ and at about 3570 and 3586 eV for UO$_2$ were 
earlier assigned to multiple-scattering resonances) 
\cite{B13,B14,B15}.  While for UO$_2$(NO$_3$)$_2\cdot$6H$_2$O, the 
structure observed at about 4 eV above the main absorption maximum was 
suggested to represent a charge-transfer satellite \cite{B13} (other 
structures at about 10 eV and 32 eV above the main absorption maximum 
were attributed to multiple scattering resonances), for UO$_2$ and 
UF$_4$, where charge-transfer effects are less pronounced, an 
identification of possible charge-transfer satellites is hampered due 
to the substantial smearing out of the spectral structures.  In this 
situation, the virtually unlimited resolution (defined by the responce 
function of the instrument) of the RIXS technique and its ability to 
enhance transitions to charge-transfer excited states are especially 
useful.

\begin{figure} [H]
\centerline{\epsfxsize=8.0cm \epsfbox{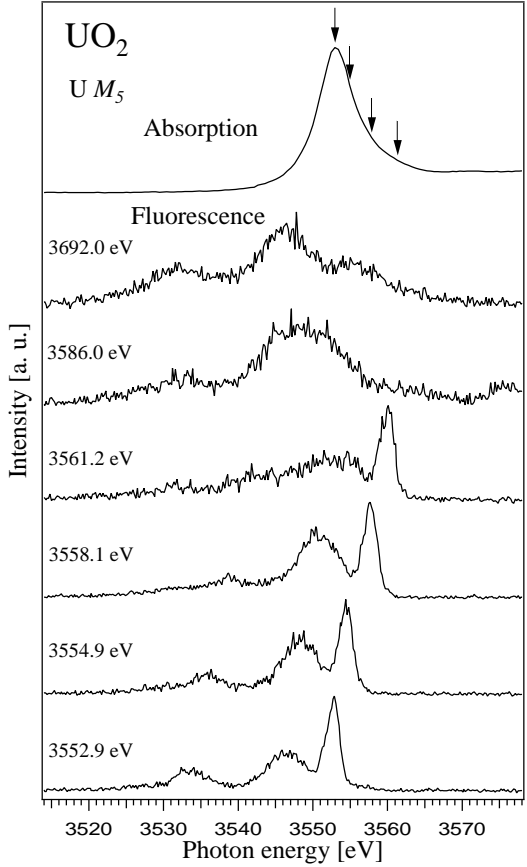}}
\caption{The total electron yield spectrum of UO$_{2}$ near the U 
$M_{5}$ absorption edge and resonant U $5f\rightarrow3d$ x-ray 
fluorescence spectra.  The arrows on the total electron yield spectrum 
indicate the excitation energies used to collect the fluorescence 
spectra.}
\label{fig18}
\end{figure}

The U 5$f\rightarrow3d$ x-ray fluorescence spectra of UO$_2$ 
(7$p\rightarrow3d$ transition probability is much lower) detected in 
the horizontal plane at 90$^\circ$ angle between directions of 
incident and scattered photons and for different excitation energies 
across the U $M_5$ absorption edge are displayed in Fig.~\ref{fig18}.  
One can identify contributions from scattering and normal fluorescence 
in these spectra.  The scattering part follows varying excitation 
energies while the normal fluorescence part appears at constant 
emitted-photon energies.  For excitation energies set near the U 
3$d_{5/2}$ threshold, the spectra consist of the ``recombination'' 
line and low-energy structure, extending over 10 eV (structures 
present at about 19 eV below the recombination peak correspond to U 
6$p_{3/2}\rightarrow3d_{5/2}$ transitions).

The shape of resonant spectra can not be attributed solely to the 
5$f^2\rightarrow3d^95f^3\rightarrow5f^2$ excitation-deexcitation 
process.  The 3$d^95f^3\rightarrow5f^2$ multiplet spread is about 4 eV 
while the separation between centroids of the ``recombination'' line 
and low-energy structure is approximately 6.5 eV. Due to significant U 
5$f-$O 2$p$ hybridization, the ground state of UO$_2$ can be described 
as a mixture of primarily 5$f^2$ and 5$f^3\underline{L}$ 
configurations.  Then, the intermediate state of the spectroscopic 
process is mainly a mixture of 3$d^95f^3$ and 3$d^95f^4\underline{L}$ 
configurations so that there is a radiative decay to 5$f^2$ and 
5$f^3\underline{L}$ states, i.~e.  transitions back to the ground 
state and to low-lying excited states.  Final states of this second 
order optical process can be divided into three categories: bonding 
(the ``recombination'' line), nonbonding, and antibonding (the 
low-energy structure) states between 5$f^2$ and 5$f^3\underline{L}$ 
configurations.  The whole low-energy structure grows slightly and 
then decreases with increasing excitation energies, showing a hint of 
some enhancement at about 8 eV below the ``recombination'' line for 
3558.1-eV incident photons. 

The energy separation between the slightly resonating structure and 
the ``recombination'' line is close to that between the 8.5-eV 
satellite and 1-eV main line in resonant valence band photoemission 
spectra of UO$_2$ (Ref.~\cite{B16}) which have been attributed to 
antibonding and bonding states between 5$f^1$ and 5$f^2\underline{L}$ 
configurations, respectively \cite{B7}.  Taking into account the 
configuration dependent hybridization \cite{R3}, the '8-eV' 
enhancement in the scattering spectrum can be tentatively associated 
with transitions to antibonding states between 5$f^2$ and 
5$f^3\underline{L}$ configurations.  Setting $V$ to 1.3 eV (due to 
higher localization of $5f$ wavefunctions, bare $V$ in this oxide is 
expected to be lower than the one derived for 
UO$_2$(NO$_3$)$_2\cdot$6H$_2$O, see below) we then estimate $\Delta$ 
($=\varepsilon_f-\varepsilon_p$) in UO$_2$ to be about 6 eV. This 
value is larger than 4.5 eV derived for the on-site $f$-$f$ Coulomb 
interaction U$_{ff}$ from photoemis-

\noindent sion and bremsstrahlung isochromat 
data \cite{B8,B17}, thus suggesting that UO$_2$ is a Mott-Hubbard-type 
insulator \cite{R2}.  This conclusion is consistent with those made 
from the analysis of both resonant and non-resonant $4f$ photoemission 
spectra \cite{B7,B8}.

\begin{figure}
\centerline{\epsfxsize=6.75cm \epsfbox{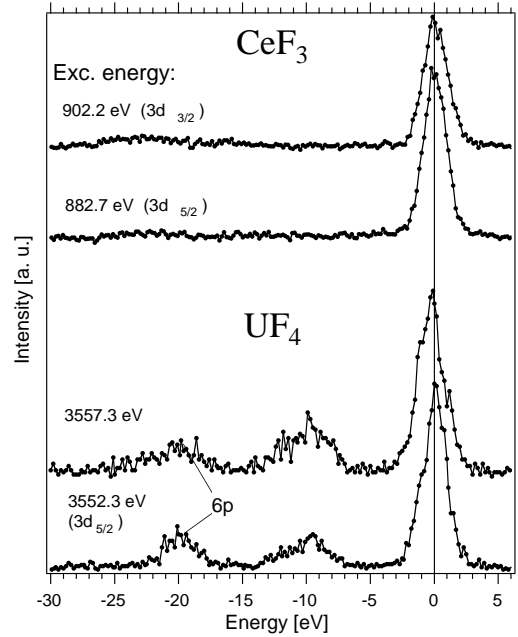}}
\caption{Resonant x-ray scattering spectra of UF$_{4}$ recorded at 
excitation energies set to the main U $M_{5}$ absorption maximum and 
to 5 eV above it together with those of CeF$_3$ obtained at both Ce 
3$d_{5/2}$ and 3$d_{3/2}$ thresholds (Ref.~[10]).}
\label{fig19}
\end{figure}

The low-energy structure that corresponds to transitions involving 
ligand 2$p\rightarrow$U 5$f$ charge-transfer excitations is also 
present in the RIXS spectra of UF$_4$ (Fig.~\ref{fig19}) obtained at 
the excitation energy set to the U $M_5$ absorption maximum and to 
5~eV above it.  The situation for UF$_4$ contrasts to that for CeF$_3$ 
where charge-transfer effects are not observed in the RIXS data 
recorded at Ce $3d$ thresholds.  The shape of the UF$_4$ spectra in 
Fig.~\ref{fig19} thus concurs with the common expectation that 5$f$ 
states in actinides are more extended than 4$f$ states in lanthanides.  
The relative intensity of the charge-transfer structure in the RIXS 
spectra of UF$_4$ is lower than that in UO$_2$ whereas the energy 
separation between centroids of this structure and the 
``recombination'' line, which is mainly determined by values of 
$\Delta$ and $V$ in the ground state, increases up to $\sim$9 eV. 
Since $V$ in UF$_4$ is rather close to that in UO$_2$, an increase 
of this energy separation observed for uranium fluoride is mainly due 
to an increase of $\Delta$.

Relative intensities of structures in scattering spectra depend not 
only on values of model parameters in the ground state of the system 
but also on the channel interference and relationship between on-site 
$f$-$f$ Coulomb interaction $U_{ff}$ and the attractive core-hole 
potential $U_{cf}$ in the intermediate state of the spectroscopic 
process.  For the spectral intensity, given by formula~(2), the 
interference effects lead to an enhancement or suppression of 
transitions to a certain final state due to a summation over $i$ 
inside the modulus.  Since $U_{cf}$ is usually somewhat larger than 
$U_{ff}$, the effective charge-transfer energy $\Delta_{\mathrm 
{eff}}=\Delta-U_{cf}+U_{ff}$ decreases in the intermediate state, thus 
giving rise to stronger configuration mixing.  On the other hand, $V$ 
is shown to be reduced in the presence of a core hole \cite{R3}.  This 
leads to a weakening of the U 5$f-$ligand 2$p$ hybridization.  In 
addition, the relative intensities of scattering structures may be 
affected by self-absorption.  Therefore, the difference in relative 
intensities of the ``recombination'' line and the charge-transfer 
structure between UO$_2$ and UF$_4$ may not necessarily be determined 
by the difference in the 5$f^3\underline{L}$ admixture in the ground 
state of these compounds.

\begin{figure}
\centerline{\epsfxsize=7.25cm \epsfbox{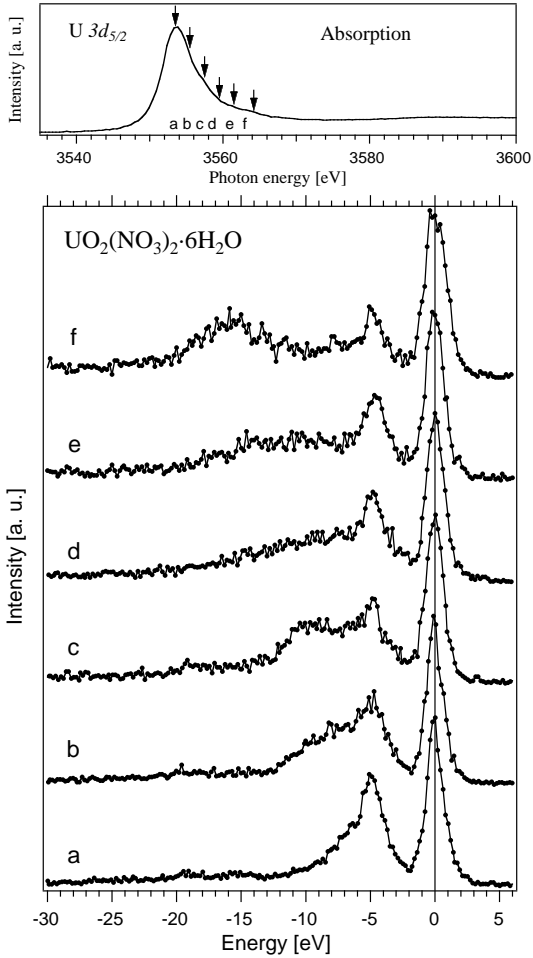}}
\caption{Resonant x-ray scattering spectra of 
UO$_2$(NO$_3$)$_2\cdot$6H$_2$O recorded at various energies of 
incident photons near the U 3$d_{5/2}$ threshold.  The arrows on the 
total electron yield spectrum of the U $M_{5}$ absorption edge, shown 
in the top panel, indicate the excitation energies used for scattering 
measurements.}
\label{fig20}
\end{figure}

In UO$_2$(NO$_3$)$_2\cdot$6H$_2$O, a degree of covalent character of 
the chemical bonds is high because U is in the highest oxidation state 
6+.  The admixture of the 5$f^2\underline{L}^2$ configuration is 
expected to be significant so that the ground state of the system can 
be described as a mixture of 5$f^0$, 5$f^1\underline{L}$, and 
5$f^2\underline{L}^2$ configurations.  As a consequence, there are 
large changes in the shape of resonant x-ray scattering spectra with 
varying excitation energies (Fig.~\ref{fig20}).  For the excitation 
energy set to the U 3$d_{5/2}$ absorption satellite at 3557.5 eV, one 
can observe an enhancement in the inelastic scattering weight at an 
energy loss of about 9.5 eV (spectrum $c$).  Similar enhancement was 
detected in the RIXS spectra of UO$_3$ (Fig.~\ref{fig2}) at 
corresponding excitation energies although, in the present case, this 
resonance is more pronounced.  The resonance indicates the 
charge-transfer character of the absorption satellite at about 4 eV 
above the U 3$d_{5/2}$ maximum.  Referring to the discussion for 
UO$_3$ (Ref.~\cite{R5}), the elastic peak and structures with energy 
losses of $\sim$5.1 eV and $\sim$9.5 eV in the scattering spectra of 
UO$_2$(NO$_3$)$_2\cdot$6H$_2$O can be associated with transitions to 
bonding, nonbonding, and antibonding states between 5$f^0$ and 
5$f^1\underline{L}$ configurations, respectively.

For UO$_3$ (Ref.~\cite{R5}), a resonance of transitions to nonbonding 
states between 5$f^1\underline{L}$ and 5$f^2\underline{L}^2$ 
configurations was also observed in scattering spectra at an energy 
loss of about 14.5 eV when the excitation energy was set to the U 
3$d_{5/2}$ absorption satellite $\sim$10 eV above the main maximum 
(see also Fig.~\ref{fig2}).  This in turn supported the assignment of 
the latter satellite to the one originating from the O 
2$p\rightarrow$U 5$f$ charge-transfer.  For 
UO$_2$(NO$_3$)$_2\cdot$6H$_2$O, spectrum $f$ recorded at similar 
excitation energy (3564.2 eV versus 3563.9 eV for UO$_3$) exhibits a 
broad line with energy losess of around 15.4 eV. However, the origin 
of this line is not clear because of uncertainty in the energy of 
normal fluorescence transitions and their relative contribution to 
spectrum $f$ (unfortunately, the high-energy excited spectra, where 
normal fluorescence dominates, were not recorded).  The broad line can 
belong to normal fluorescence, or it can correspond to a resonance of 
charge-transfer excited states as a result of coupling between 
5$f^1\underline{L}$ and 5$f^2\underline{L}^2$ configurations.  The 
possibility of some contribution of transitions to the latter states 
is suggested by the shape of other spectra of 
UO$_2$(NO$_3$)$_2\cdot$6H$_2$O recorded at lower excitation energies.  
For example, spectra $c$ and $e$ (Fig.~\ref{fig20}) contain structures 
with similar energy losses to those for the broad line in spectrum 
$f$.

Some differences in the behavior of RIXS spectra between 
UO$_2$(NO$_3$)$_2\cdot$6H$_2$O and UO$_3$ which both contain U$^{6+}$ 
are due to somewhat different environment for U in these compounds.  
In UO$_2$(NO$_3$)$_2\cdot$6H$_2$O, the U ion is surrounded by eight O 
ions \cite{B18} which create two ``short'' and six ``long'' U--O bonds 
of 1.76 {\AA} and 2.48 {\AA}, respectively.  In UO$_3$, U has six 
nearest O neighbors \cite{B19} with two of them located at the 
1.79-{\AA} distance and others at 2.30 {\AA}.  This strong 
inequivalence of O sites implies a large variation in the value of $V$ 
for the same compound since $V$ is expected to scale with the 
cation-anion distance.  The values of $\Delta$ may also be different 
for inequivalent U--O bonds.

\begin{figure}
\centerline{\epsfxsize=7.0cm \epsfbox{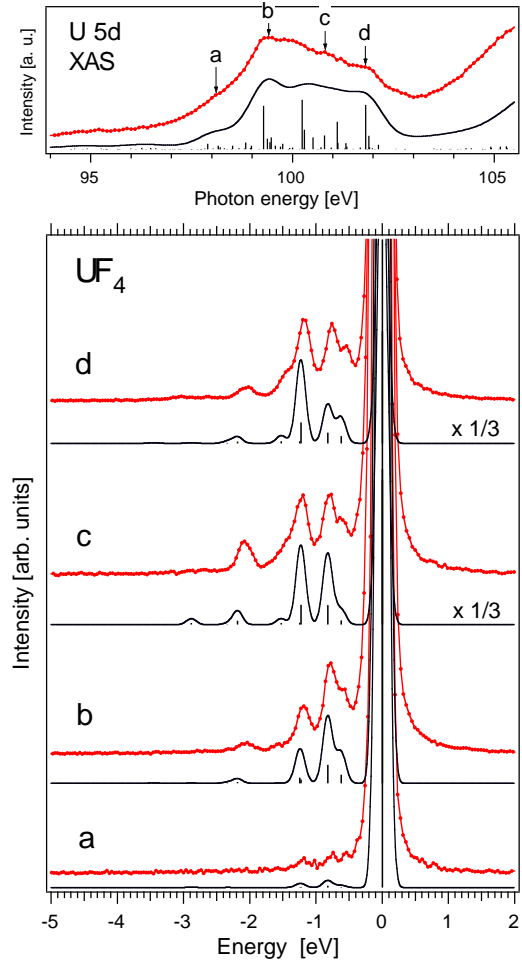}}
\caption{Resonant x-ray scattering spectra of UF$_4$ (Ref.~\cite{B20}) 
recorded at different excitation energies close to the U 5$d$ 
threshold (lines with markers) together with the results of atomic 
multiplet calculations (sticks with thin lines) for the U$^{4+}$ ion.  
Excitation energies are indicated by arrows on the total electron 
yield spectrum at the U $5d$ absorption edge shown in the top panel.}
\label{fig21}
\end{figure}

The determination of the energies of transitions to bonding, 
nonbonding and antibonding states between 5$f^0$, 5$f^1\underline{L}$, 
and 5$f^2\underline{L}^2$ configurations from resonances in scattering 
spectra puts additional constrains on values of $V$, $\Delta$, and 
$U_{ff}$ in the ground state of UO$_2$(NO$_3$)$_2\cdot$6H$_2$O. 
Neglecting the inequivalence of U-O bonds, model parameters can be 
estimated by diagonalizing a simplified Hamiltonian so that its 
eigenvalues coincide with energies of corresponding states.  This 
gives 1.4, 3.5, and 4 eV for $V$, $\Delta$, and $U_{ff}$, 
respectively.  The derived average values suggest that 
UO$_2$(NO$_3$)$_2\cdot$6H$_2$O is in the intermediate regime of the 
Zaanen-Sawatzky-Allen diagram \cite{R2}.

RIXS measurements at the actinide $5d$ threshold provide an 
opportunity to study in detail elementary excitations in actinide 
compounds due to the naturally higher resolution of such experiments 
in comparison with those at the actinide $3d$ and $4d$ thresholds.  An 
example of probing the $f$-$f$ excitations in actinide systems is 
illustrated in Fig.~\ref{fig21} where the RIXS spectra of solid 
UF$_4$, recorded for different incident photon energies in the 
pre-$5d$-threshold region, are displayed.  The assignment of sharp 
inelastic scattering structures to the $f$-$f$ transitions is 
supported by atomic multiplet calculations for the U$^{4+}$ ion.  The 
spectra were calculated using equation~(2), where the varying lifetime 
of core-excited states due to the autoionization via the $5d$-$5f5f$ 
super Coster-Kronig decay was taken into account.  The autoionization 
into the continuum of $g$ symmetry was only considered since it is the 
most dominant path.  Matrix elements were obtained from Cowan's 
programs so that Slater integrals $F^k(5f,5f)$, $F^k(5d,5f)$, 
$G^k(5d,5f)$, and $R^k(5d\epsilon{g},5f)$ were scaled down to 75\%, 
75\%, 66\%, and 80\%, respectively, from the Hartree-Fock values.  The 
density of states of the continuum was assumed to be constant and the 
kinetic energy of the continuum electron was set to the value which 
made the average energies of $5d^95f^3$ and $5d^{10}5f^1\epsilon{g}$ 
equal.

The calculations reproduce all of the spectral structures very well 
especially an enhancement of the peak at about 1.2 eV with increasing 
excitation energies.  The growth of the peak is due to enhanced 
transitions into the $^1G_4$ state.  Changes in absolute intensities 
of inelastic scattering structures corresponding to the $f$-$f$ 
transitions are reproduced on going from spectrum $a$ to spectrum $b$.  
For spectra $c$ and $d$, such changes in calculated intensities are 
about three times higher as compared to those in the experiment.  The 
discrepancy may originate from the normalization procedure for the 
experimental spectra to account for variations in the incident photon 
flux.  The intensity of the elastic peak was used as a reference in 
this procedure.  However, the elastic peak contains some contribution 
of diffuse scattering which may vary with varying excitation energies.

RIXS profiles, corresponding to the $f$-$f$ excitations, are found to 
be very sensitive to the chemical state of U in different systems 
\cite{B20}.  For example, it is a matter of the presence or absence of 
these excitations when going from U$^{4+}$ to U$^{6+}$ compounds.  
Even for the same oxidation state, the corresponding RIXS structures 
are observed to be broadened in compounds with the increasing degree 
of covalency in chemical bonding.  Therefore, RIXS measurements near 
the U $5d$ threshold provide good fingerprints for the chemical state 
of U in different systems in contrast to x-ray absorption spectra 
which show only small differences at the U $5d$ edge \cite{B14,B20}.

\paragraph{ Acknowledgements}
Much of experimental work reviewed here have been carried out in 
collaboration with Prof.  Joseph Nordgren, Dr.  Akane Agui, Dr.  
Laurent Duda, Dr.  Jinghua Guo, Dr.  Pieter Kuiper, Dr.  Yanjun Ma, 
Dr.  Martin Magnuson, Dr.  Derrick Mancini, Dr.  Ken Miyano, Dr.  
David Shuh, and Conny S{\aa}the.  The author is also grateful to Dr.  
Frank de Groot for making available TT-MULTIPLETS programs 
and for his help in using those.  This work was supported by the 
Swedish Natural Science Research Council (NFR) and the G\"{o}ran 
Gustavsson Foundation for Research in Natural Sciences and Medicine.

\end{document}